\documentclass[aps,pre,twocolumn,superscriptaddress,nofootinbib]{revtex4-2}
\usepackage{ragged2e}
\usepackage[english]{babel}
\usepackage{amsmath}
\usepackage{amsfonts}
\usepackage{amssymb}
\usepackage{graphicx}
\usepackage[utf8]{inputenc}
\usepackage{array}
\usepackage{physics}
\usepackage{xr-hyper}
\usepackage{hyperref}
\usepackage{xcolor}
\usepackage[normalem]{ulem}
\usepackage{booktabs}

\begin{document}
\title{Resonance Assisted Tunneling in Floquet spin-J systems.}
\author{J. A. Segura-Landa}
\affiliation{Instituto de Ciencias Nucleares, Universidad Nacional Aut\'onoma de M\'exico, Apdo. Postal 70-543, C.P. 04510  Cd. Mx., Mexico}

\author{Diego A. Wisniacki}
\affiliation{Departamento de F\'isica ``J. J. Giambiagi'' and IFIBA, FCEyN,
Universidad de Buenos Aires, 1428 Buenos Aires, Argentina}

\author{Sergio Lerma-Hernández}
\affiliation{Facultad  de F\'isica, Universidad Veracruzana,  Campus Arco Sur, Paseo 112, C.P. 91097  Xalapa, Mexico.}

\author{Jorge G. Hirsch}
\affiliation{Instituto de Ciencias Nucleares, Universidad Nacional Aut\'onoma de M\'exico, Apdo. Postal 70-543, C.P. 04510  Cd. Mx., Mexico}

\begin{abstract}
   We apply the theory of Resonance Assisted Tunneling (RAT) to a many-body quantum kicked system with a well-defined semiclassical limit. Using a quantum resonant condition, we identify eigenstates associated with classical resonances and compute their quasienergy splitting semiclassically. We distinguish two regimes: the first, where RAT predictions show excellent agreement with exact quantum results, and a second, where the splitting saturates and coincides with that of a harmonic oscillator with frequency determined by the classical oscillation of the resonance island. We quantify the perturbation strength above which RAT theory is no longer valid and analyze its scaling in the semiclassical limit, providing analytical expressions to estimate this upper bound. 

\end{abstract}
\maketitle
\section{Introduction}
Many-body quantum systems play a central role in modern physics, both for the fundamental insights they provide into collective quantum phenomena \cite{fetter2003quantum}, and for their potential technological applications, including quantum simulation \cite{QuantumSimulation} and quantum information processing \cite{bruss2019quantum}. Advances in experimental platforms have turned their study from a mainly theoretical challenge into a practical and highly tunable reality. Ultracold atoms in optical lattices \cite{Bloch2012,UltraColdAtoms}, trapped ions \cite{Blatt2012,RevModPhys.93.025001}, superconducting qubits \cite{Houck2012}, and photonic lattices \cite{Rechtsman2013} allow researchers to implement Hamiltonians with remarkable precision, controlling interaction strengths, dimensionality, and external driving. This capability has enabled the exploration of systems in broad range of parameters and their comparison with many-body theoretical predictions.

Within this broad variety of systems, there exists a subset that admits a well-defined semiclassical limit. In this limit, where the effective Planck constant tends to zero $\hbar_\text{eff}\to0$, quantum mechanics gives rise to an effective description in terms of a classical phase space \cite{Brack2003,RevModPhys.54.407,POLKOVNIKOV20101790}, in which one can analyze the influence of various classical structures on the quantum regime. The use of tools from classical mechanics further expands our understanding of such systems, allowing us to anticipate phenomena such as quantum phase transitions \cite{RevModPhys.69.315,MatthiasVojta_2003}. 

A particular example of the impact that classical structures can have on the quantum regime are nonlinear resonances in the underlying classical phase space. In mixed systems, these resonances, which are characterized by two integers $r{:}s$, appear as chains of periodic orbits, or “islands”, which mediate the coupling between quantum states localized in distinct regular regions. This coupling can bridge states that are energetically far apart and would otherwise be only weakly connected through direct tunneling. The mechanism responsible for this enhancement is known as resonance-assisted tunneling (RAT) \cite{BRODIER200288}, a semiclassical process that has been extensively analyzed in paradigmatic models such as the Standard and Harper maps \cite{BRODIER200288,PRLRAT,Wisniacki_2011,Wisniacki_2014,Wisniacki_2015},  where RAT theoretical predictions show excellent agreement with numerical results. More recently,  RAT has been  experimentally observed in asymmetrically deformed microcavities (ADM's) \cite{Kwak2015} and Periodically-driven Bose-Einstein condensates \cite{PDBoseEinstein}.

In this contribution, we apply RAT theory to the resonant states of a periodically kicked multi-spin system. Since the perturbation period can be externally controlled, we can study distinct resonances within the same energetic region in our system. We idenfity two well-defined regimes. The first regime occurs when the system is weakly perturbed, and the classical resonance structures are smaller in phase space than the effective Planck constant. In this case, RAT theory shows good agreement with the exact quantum results. The second regime occurs when the perturbation is large enough that the area of the resonance islands become comparable to, or larger than, the effective Planck constant. In this situation, RAT theory loses accuracy. We quantify this upper bound on the perturbation strength and show that it coincides with the point where the states become Einstein–Brillouin–Keller (EBK) quantized within the islands, thus becoming eigenstates of an effective harmonic oscillator with a frequency given by the local oscillation frequency of the resonance. Finally, we show that this bound tends to zero in the semiclassical limit, following a power-law behavior that  we explicitly  determine.

This paper is organized as follows. In Sec.~II we introduce the quantum system under study and present its semiclassical Hamiltonian. Section~III establishes the condition for a classical $r{:}s$ resonance, which is subsequently extended to the quantum regime, leading to the definition of resonant states. We also show how the splitting of eigenphases between resonant states is  estimated semiclassically. Section~IV contains the main results of this work, where the two regimes are identified and the scaling behavior of the maximum perturbation strength is derived. Finally, our conclusions are summarized in Sec.~V.

\begin{figure*}
    \centering
    \includegraphics[width=\linewidth]{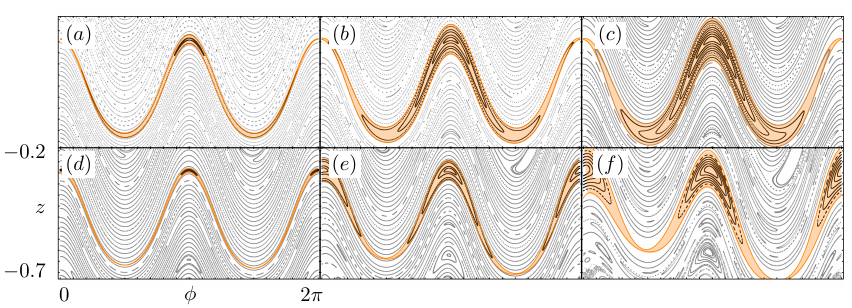}
    \caption{Poincaré sections of classical resonances. Panel (a)-(c) shows the growth of a resonance island for a $1{:}1$ resonance with $\tau=8$ for $(a) \,\,\epsilon=0.0005$, $(b) \,\,\epsilon=0.005$ and $(c) \,\,\epsilon=0.01$. Panel (d)-(f) shows the growth of a resonance island for a $2{:}1$ resonance with $\tau=4$ for $(d) \,\,\epsilon=0.002$, $(e) \,\,\epsilon=0.1$ and $(f) \,\,\epsilon=0.2$. In all cases, the phase space area $S_{r:s}^+-S_{r:s}^-$ is the region highlighted in orange.}
    \label{fig: Poincare sections}
\end{figure*}
\section{The Model}
\subsection{Quantum Hamiltonian}
We consider a time-dependent spin-$J$ system described by the Hamiltonian  
\begin{equation}
    \hat{H}(t) = \hat{H}_0(\vec{J}) + \epsilon\,\hat{K}(\vec{J})\sum_{n=-\infty}^{\infty}\delta(t-n\tau),
\end{equation}  
where $\vec{J} = (\hat{J}_x, \hat{J}_y, \hat{J}_z)$ are collective spin operators satisfying the SU$(2)$ algebra, $[\hat{J}_i, \hat{J}_j] = i \epsilon_{ijk} \hat{J}_k$ ($\hbar = 1$). The operator $\hat{H}_0$ is chosen  such that its  classical limit corresponds to an integrable  Hamiltonian with   one degree of freedom. In turn, the $\tau$-periodic perturbation $\hat{K}(\vec{J})$ introduces non-integrability through the dimensionless perturbation strength $\epsilon$. Hamiltonians of this form have been extensively investigated in diverse contexts, from quantum chaos \cite{Haake1987,Ghose2008PRA,Dogra2019PRE} to quantum simulation \cite{sieberer2019digital,ManuelPRL2020,Chinni2022}, and have also been experimentally realized~\cite{Chaudhury2009nature,Neill2016,Krithika2019PRE}. The spectral and dynamical properties can be studied using Floquet theory\cite{floquet1883equations,GRIFONI1998229}, with the stroboscopic evolution given by the Floquet operator
\begin{equation}
    \hat{F} = \exp\left(-i \tau \hat{H}_0 \right) \exp\left(-i \epsilon \hat{K} \right),
\end{equation}
whose eigenstates \( \ket{f_k} \) satisfy
\begin{equation}
    \hat{F} \ket{f_k} = e^{-i\varphi_k} \ket{f_k},
\end{equation}
with quasienergies \( \varphi_k \) playing a role analogous to energy eigenvalues in time-independent systems.

In our case, the system corresponds to a kicked version of the Lipkin–Meshkov–Glick (LMG) model, which describes $N = 2J$ mutually interacting spins in a static magnetic field along the $z$-axis. In the pseudo-spin representation, the static part of the Hamiltonian is given by \cite{LIPKIN1,LIPKIN2,LIPKIN3} 
\begin{equation}
    \label{eq: H0}
    \hat{H}_0 = \omega_0 \hat{J}_z + \frac{\gamma_x}{2J - 1} \hat{J}_x^2,
\end{equation}  
where $\omega_0$ denotes the Larmor frequency associated to a field in the $z$-direction and $\gamma_x$ characterizes the nonlinear inter-spins interactions along $x$-direction. The periodic driving is implemented by choosing $\hat{K} = \hat{J}_x$, which can be regarded as an additional magnetic field along the $x$-axis applied periodically with period $\tau$ and amplitude controlled by the parameter $\epsilon$. 

The Hamiltonian \eqref{eq: H0} commutes with the two-eigenvalued ($\pm 1$) parity operator $\hat{\Pi}=e^{i \pi(\hat{J_z}-J)}$, where $J$ is the angular momentum quantum number. We refer to the eigenstates $\ket{E_k}$ with $\hat{H}_0\ket{E_k}=E_k\ket{E_k}$ as integrable (unperturbed) states. There are a total of $2J+1$ of such states, which can be classified into the two parity sectors.  Unless otherwise stated, we set \( \omega_0 = 1 \) and \( \gamma_x = -0.95 \) throughout, which corresponds to the region I of the LMG model previously reported in \cite{Ribeiro,ribeiro2008exact,NaderAvoided}. In this parameter regime, the energy spectrum of $\hat{H}_0$ consists of non-degenerate levels, with  parity eigenvalues alternating  from one level to the next.

In the next section we will show how by manipulating adequately the kicking period $\tau$, we can couple pairs of integrable states in distinct energetic regions. In particular, we couple quantum states whose quantum number differs in multiples of $r$. The effective Planck constant is given as $\hbar_{\text{eff}}=1/J$ \cite{Ribeiro2006}. Hence, the semiclassical limit is reached as $J\to\infty$, which in this case is equivalent to the thermodynamic limit of the system. 
\subsection{Classical Hamiltonian}
A classical counterpart of the quantum Hamiltonian \eqref{eq: H0} is derived by computing its expectation value respect to Bloch coherent states, defined as 
\begin{equation}
    \ket{\alpha}=\frac{1}{(1+\abs{\alpha}^2)^J}\exp\left(\alpha\hat{J}_+\right)\ket{J\,,-J},
\end{equation}
where $\alpha$ is a complex parameter $\alpha=\tan(\theta/2)e^{-i\phi}$ in the unitary Bloch sphere. By taking the limit  
\begin{equation}
    H_0=\lim_{J\to\infty}\frac{1}{J}\mel{\alpha}{\hat{H}_0}{\alpha},
\end{equation}
we obtain the classical Hamiltonian:
\begin{equation}
    \label{eq: classical hamiltonian}
    \begin{split}
        H_0(\phi, z) = &\omega_0 z + \left(\frac{1 - z^2}{2}\right)\gamma_x \cos^2 \phi 
    \end{split}
\end{equation}
where \( z = -\cos\theta \) and \( \phi \) are canonical conjugate variables describing the motion on the Bloch sphere. The classical system has a single degree of freedom and the energy is conserved, hence it is classically integrable. This integrability is broken when the full-time dependent system is considered.

Contrary to other kicked-top systems \cite{Haake1987}, for the present system there is no an analytical mapping describing the full time-dependent Hamiltonian stroboscopic dynamics. Instead, the stroboscopic dynamics is obtained numerically as follows. In the time intervals $t\in[(n-1)\tau,n\tau]$ for the $n-$th iteration,  the classical orbits of $H_0$ are numerically computed from the Hamilton equations 
\begin{equation}
    \begin{split}\label{eq: HamEq}
        \dot{\phi}=\pdv{H_0}{z}, \quad \dot{z}=-\pdv{H_0}{\phi}.
    \end{split}
\end{equation}
 Then, the periodic driving is added as an instantaneous rotation, by an angle $\epsilon$  around the $x$-axis,   at $t=n\tau$. For large $\epsilon$, classical trajectories may exhibit chaotic behavior. However, in the present work, we restrict our attention to the quasi-integrable regime \( \epsilon \ll 1 \), where perturbative and semiclassical techniques are applicable.

\section{Non-linear resonances}

\subsection{Classical resonance condition}
Classically, an $r{:}s$ resonance in a periodically kicked one-degree-of-freedom system with period $\tau$ occurs whenever the condition
\begin{equation}\label{eq: classical resonance condition}
    r\tau = sT(\mathcal{E}), \quad r, s \in \mathbb{Z},
\end{equation}
is satisfied. Here, $T(\mathcal{E})$ is the oscillation period of the classical orbit of $H_0$ with classical energy $\mathcal{E}=E/J$, which for our case is calculated as 
\begin{equation}
    T(\mathcal{E}) = \int\limits_{0}^{2\pi} \frac{d\phi}{\sqrt{1 - 2\mathcal{E}\,\Gamma(\phi) + \Gamma(\phi)^2}},
\end{equation}
where \( \Gamma(\phi) = \gamma_x \cos^2 \phi \). It is worth noting that, since the driving period $\tau$ can be externally controlled, different resonances can be explored using the same orbit of $H_0$. 

Figure~\ref{fig: Poincare sections} illustrates this methodology through stroboscopic Poincaré sections of the system for distinct values of $\tau$. We begin by identifying the orbit of $H_0$ whose oscillation period is $T(\mathcal{E}) = 8$. We refer to the energy of this orbit as the \emph{resonant energy}, denoted by $\mathcal{E}_R$, which for the parameters used in this work is $\mathcal{E}_R =-0.723276$. Panels~(a)–(c) display a 1-chain of islands associated with a $1{:}1$ resonance, obtained by setting $\tau = 8$. In this case, the classical resonant orbit with energy $\mathcal{E}_R$ generates a $1{:}1$ resonance, as stated by the Poincaré–Birkhoff theorem. Note that the growth of the resonance can be further controlled through the perturbation strength $\epsilon$. Fig.~\ref{fig: Poincare sections} Panel~(a) shows the resonance for $\epsilon = 0.0005$, where the phase-space area it occupies is relatively small. In Panel~(b), with $\epsilon = 0.005$, the resonance island has grown noticeably, and finally, in Panel~(c), with $\epsilon =0.01$, the island reaches a considerable size.

Fig.~\ref{fig: Poincare sections} Panels~(d)–(f) show the analogous situation for $\tau = 4$, where the same classical tori with energy $\mathcal{E}_R$ now give rise to a $2{:}1$ resonance, since for this case $T(\mathcal{E}_R) = 2\tau$. Fig.~\ref{fig: Poincare sections} Panel~(d) shows the resonance for $\epsilon = 0.02$, Panel (d) for $\epsilon=0.1$, and Panel (f) for $\epsilon=0.2$. Note that the $2{:}1$ resonance requires  much larger values of   $\epsilon$ to manifest. This differentiated sensitivity to the perturbation  is closely related with the symmetries of the classical Hamiltonian ${H}_0$, which is invariant under the transformations $\phi \rightarrow -\phi$  and $\phi \rightarrow \phi + \pi$. The perturbation leading to the 2:1 resonance breaks the second symmetry. It has two effects which can be clearly seen comparing the panels (c) and (f) of Fig.~\ref{fig: Poincare sections}. The 1:1 resonance has only one stable fixed point located at $\phi_f= \pi$, while the 2:1 resonance has two stable  fixed points, located at $\phi_{f1}= 0$ and $\phi_{f1}= \pi$. It is more interesting to observe that the 2:1 has broken the symmetry around  $\phi= \pi$, exhibiting a noticeable asymmetry. Both effects combined imply a  differentiated sensitivity to the perturbation, making more difficult the growing of the resonant area for the 2:1 resonance. This differentiated sensitivity to the perturbation can also be understood form the quantum version of the model, as discussed in more detail below. 

In general, once $\mathcal{E}_R$ is chosen, a $r{:}s$ resonance is obtained by setting $\tau = s\,T(\mathcal{E}_R)/r$. However, in this work we focus on the illustrative $1{:}1$ and $2{:}1$ cases. This approach offers a quantum-mechanical advantage, as it allows us to target a specific energetic region of the spectrum, rather than searching for eigenstates that best satisfy the resonance condition for a fixed $\tau$. We discuss it in the next subsection.

\subsection{Quantum resonance condition and resonant states}
\begin{figure}
    \centering
    \includegraphics[width=0.95\linewidth]{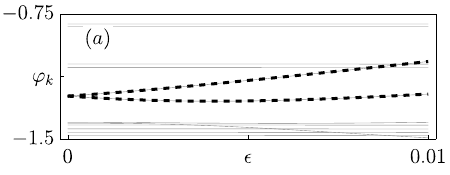}
    \includegraphics[width=0.95\linewidth]{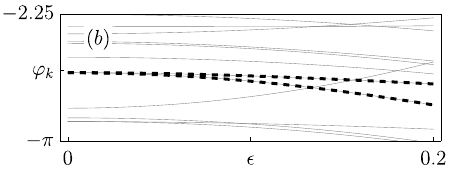}
    \caption{Sample of quasi-energies as a function of  perturbation strength for $J=30$. Panel (a): The two  quasi-energies levels associated with the $1{:}1$ resonance are highlighted using thick dashed lines. These states are degenerated at $\epsilon=0$. For $\epsilon>0$, the degeneracy  is lifted, and the resulting quasi-energy   splitting is analyzed and compared with semiclassical approximations in Sec.\ref{sec:SCA}. Panel (b): analog situation for the $2{:}1$ resonance.}
    \label{fig: quasienergies}
\end{figure}
A quantum equivalent of the classical resonance condition \eqref{eq: classical resonance condition} was introduced in \cite{segura2024quantum,seguralanda2025} as
\begin{equation}
    \label{eq: quantum resonance condition}
    \tau = s\,T_{k,r},
\end{equation}
where $T_{k,r}$ is the quantum period defined as the inverse of the transition frequency among the $k$ and $k+r$ eigenstates of $\hat{H}_0$
\begin{equation}
    \label{eq: quantum periods}
    T_{k,r} = \frac{2\pi}{E_{k+r} - E_k}.
\end{equation}
In the semiclassical limit \( J \to \infty \), where the approximation \( E_{k+r} - E_k \approx r(E_{k+1} - E_k) \) holds, \eqref{eq: quantum resonance condition} converges to the classical equation \eqref{eq: classical resonance condition}.

Analogous to the classical picture, we can use \eqref{eq: quantum periods} to study distinct quantum resonances by setting a reference state \( E_{k_R} \) and varying $r$. In this case, the index $k_R$ denotes the quantum eigenstate of $\hat{H}_0$ whose energy $E_{k_R}/J$ is closest to the classical resonance energy $\mathcal{E}_R$. Once the index \( k_R \) is determined under this condition, a \( 1{:}1 \) quantum resonance is obtained by setting $\tau=2\pi/(E_{k_R+1}-E_{k_R})$. Without changing the reference state \( k_R \), a \( 2{:}1 \) resonance is now obtained by setting $\tau=2\pi/(E_{k_R+2}-E_{k_R})$. Since for $\epsilon=0$ the quasienergies are $\varphi_k^{(0)}=\tau E_k$, the $k_R$ and $k_R+r$ states of the Floquet operator form  a degenerate pair
\begin{equation}
    \mod\left(\abs{\varphi_{k_R}^{(0)} - \varphi_{k_R + r}^{(0)}}, 2\pi\right) = 0.
\end{equation}
We call to such states as \textit{resonant states}. As shown in Fig. \ref{fig: quasienergies}, if \( \epsilon \) increases, the states gradually break their degeneracy  due to the action of $\hat{K}$. The resulting quasienergy  splitting \( \Delta\varphi_{k_R, k_R + r} \) can be  estimated semiclassically as we discuss below.

It is worth noting that, due to the discrete nature of the quantum spectrum of $\hat{H}_0$, the quantum period \eqref{eq: quantum periods} fluctuates with \( J \) (although, as shown in \cite{seguralanda2025}, it converges to the classical period in the limit \( J \to \infty \)). Hence, for a fixed value of \( \tau \), we cannot guarantee the existence of a pair of quantum resonant states which exactly satisfies \eqref{eq: quantum periods} while varying \( J \). To ensure the presence of exactly degenerate states, we adjust the value of \( \tau \) for each \( J \) in the quantum regime (see Appendix \ref{App:QP}). Classically, the dynamics remain essentially unaffected by small variations in \( \tau \), which justifies the use of a fixed classical kicking period, even though the quantum one fluctuates slightly.

\subsection{RAT approximmation}
We now show how the quasienergy splitting of resonant states \(\Delta\varphi_{k_R,\, k_R + r}\) in an $r{:}s$ resonance can be estimated semiclassically. The first step is to obtain an effective classical Hamiltonian in the vicinity of the resonance of interest. It is well-known that the dynamics near an \( r{:}s \) resonance can be adiabatically approximated by the pendulum-like integrable Hamiltonian of the form \cite{CHIRIKOV1979263,BRODIER200288}:
\begin{equation}\label{eq: Pendulum Hamiltonian}
    H_{\text{eff}}^{(r{:}s)} = \frac{(I - I_{r{:}s})^2}{2m_{r{:}s}} + 2K_{r{:}s} \cos(r\vartheta),
\end{equation}
where $I$ is the classical action of $H_0$, and $\vartheta$ is the conjugate angle to the action. The action $I_{r:s}$ denotes the resonant tori, which in our case is the one with energy $\mathcal{E}_R$. Note that we will have a different Hamiltonian \eqref{eq: Pendulum Hamiltonian} depending on the resonance of interest. The effective mass \( m_{r{:}s}=[d²H_0/dI^2(I_{r:s})]^{-1}\) and action \( I_{r{:}s} \) parameters can be computed without requiring the explicit form of \( H_0 \), simply by extracting information from the classical phase space, using the relations \cite{BRODIER200288}:
\begin{equation}\label{eq: classical eq}
    \begin{split}
        I_{r{:}s} &= \frac{1}{4\pi} \left( S^{+}_{r{:}s} + S^{-}_{r{:}s} \right), \\
        \sqrt{2m_{r{:}s}K_{r{:}s}} &= \frac{1}{16} \left( S^{+}_{r{:}s} - S^{-}_{r{:}s} \right), \\
        \sqrt{\frac{2K_{r{:}s}}{m_{r{:}s}}} &= \frac{1}{r^2\tau} \cos^{-1} \left( \frac{\operatorname{tr} M_{r{:}s}}{2} \right),
    \end{split}
\end{equation}
where \( S^{+}_{r{:}s} \) and \( S^{-}_{r{:}s} \) are the phase-space areas enclosed by the separatrix of the \( r \)-cycle (see Fig. \ref{fig: Poincare sections}), and $M_{r{:}s}$ is the monodromy matrix of the classical equations of motion \eqref{eq: HamEq} evaluated at the stable fixed point of the cycle. This allows us to avoid seeking a canonical transformation to action-angle variables, which, as in our case, is not always possible to find in terms of simple functions. 

The next step is to quantize the effective Hamiltonian~\eqref{eq: Pendulum Hamiltonian}, which in the unperturbed basis is a diagonal matrix with eigenvalues:
\begin{equation}
    \varepsilon_n = \frac{[\hbar_\text{eff}(n + 1/2) - I_{r:s}]^2}{2m_{r:s}}, \quad n \in \mathbb{Z}.
\end{equation}
In the limit \( K_{r:s} \to 0 \), the eigenstates $\varepsilon_{k+r}$ and $\varepsilon_k$ interact via the effective matrix element \( \abs{K_{r:s}} \). Thus, the level splitting between them is $\varepsilon_{k+r}-\varepsilon_k=2\abs{K_{r:s}}$. To compare with the exact quantum results of the Floquet eigenbasis, we use the fact that $\varepsilon_k= \hbar_\text{eff}\varphi_k/\tau$ and hence
\begin{equation}\label{eq: quasienergy splitting}
    \hbar_{\text{eff}}\Delta\varphi_{k+r,k}/\tau = 2\abs{K_{r:s}},
\end{equation}
where the parameter \( K_{r:s} \) is numerically computed from the relations~\eqref{eq: classical eq}, and depends on the size of the resonance island.
\subsection{Harmonic approximation}
The Hamiltonian \eqref{eq: Pendulum Hamiltonian} can be further simplified by making a harmonic approximation of the effective potential near the energy minimum as
\begin{equation}\label{eq: Hamonic Hamiltonian}
    H_{\text{eff}}^{(r{:}s)} \approx \frac{(I - I_{r{:}s})^2}{2m_{r{:}s}} - K_{r{:}s}r^2\theta^2.
\end{equation}
Hence, at the center of each island, the dynamics is that of a harmonic oscillator with frequency given by $\omega=r\sqrt{2\abs{K_{r:s}}/m_{r:s}}$.  Once the quantum states are fully localized within the resonance islands, we can use the approximation~\eqref{eq: Hamonic Hamiltonian}, and the level splitting becomes the constant splitting of a harmonic oscillator:
\begin{equation}\label{eq: quasienergy splitting 2}
    \Delta\varphi_{k,k+r}/\tau = r \sqrt{\frac{2\abs{K_{r:s}}}{m_{r:s}}}.
\end{equation}
The Hamonic regime can be reached either by increasing the island size through a larger $\epsilon$, or by shrinking the quantum states with a smaller $\hbar_{\text{eff}}$, so that they are more localized within the islands.

\section{Semiclassical analysis}
\label{sec:SCA}
In this section, we present the core results of our study. The discussion is divided into three subsections. In the first subsection, we describe in detail the procedure employed to determine the classical parameters entering the effective Hamiltonian [Eq.~\eqref{eq: Pendulum Hamiltonian}]. These parameters are extracted from the analysis of classical Poincaré sections. The second subsection is devoted to the semiclassical approximation of the quasienergy splitting. Here, we show how the classical parameters obtained previously are incorporated into the semiclassical expression to the level splitting. In particular, we show explicitly the existence of the two regimes. Finally, in the third subsection, we focus on the transition between such regimes and provide a maximum perturbation strength, along with its scaling in the semiclassical limit. 
\subsection{Calculation of classical elements}
\begin{figure}
    \centering
    \includegraphics[width=\linewidth]{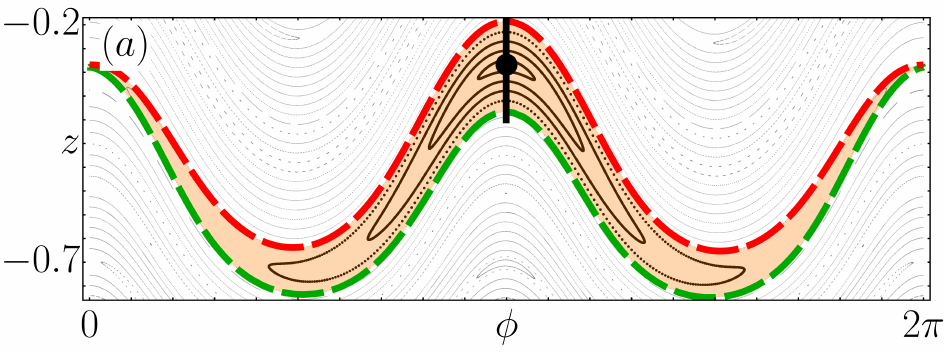}
    \includegraphics[width=\linewidth]{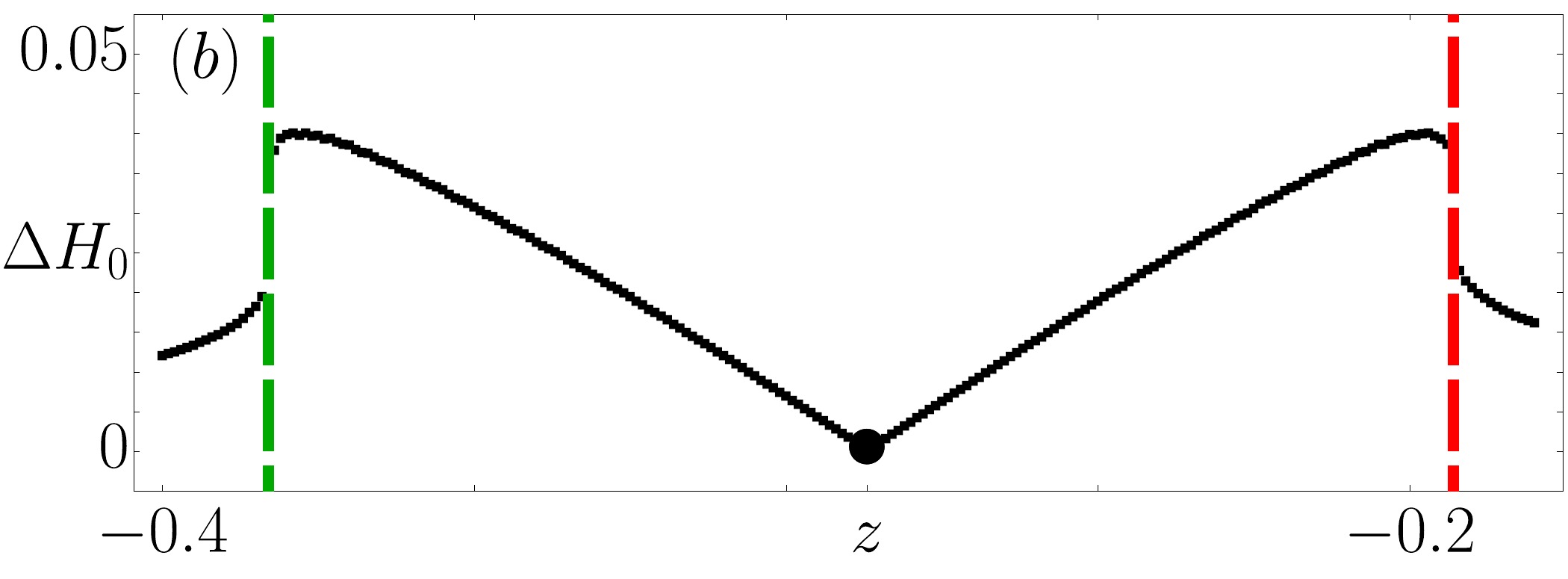}
    \caption{Panel (a): Poincaré section for a $1{:}1$ resonance with $\epsilon=0.01$. The horizontal black line indicates a set of initial conditions with $\phi=\pi$ and varying $z$, such that the island is crossed. The phase space area below the red curve is $S_{r:s}^{+}$. In turn, $S_{r:s}^{-}$ is the area below the green curve. The stable fixed point is indicated with a black circle. Panel (b): Standard deviation $\Delta{H_0}$ through the black line. Separatrices and the stable fixed point are highlighted. }
    \label{fig: appendix}
\end{figure}
In this subsection we show how we solve the system of equations \eqref{eq: classical eq} in order to find the classical parameters $K_{r:s}$, $I_{r:s}$, and $m_{r:s}$. The procedure is as follows. For a given value of $\epsilon$, the resonance island is first identified. As mentioned throughout the text, the resonance energy $\mathcal{E}_R$ is fixed, which ensures that the resonance island of interest remains in the same region of the classical phase space for both resonances.  

Once the resonance island is located, we numerically determine the associated separatrix with the highest possible precision. This is achieved by evolving the classical equations of motion for a set of initial conditions along the line $\phi=\pi/r$ while varying $z$, making sure that the chosen line intersects the island. The latter is shown in Fig. \ref{fig: appendix} Panel (a) for $r=1$, where a solid black line indicates the set of initial conditions considered, which crosses the island from the upper branch of its separatrix, passes through the stable fixed point (highlighted with a black dot), to finally cross again the separatrix in its lower branch. For every trajectory $\vec{x}(\pi,z_0,N)=((\pi,z_0),(\phi_1,z_1),\dots(\phi_N,z_N))$ generated from each initial condition in the solid black line, we compute the standard deviation of the energy
\begin{equation}
    \Delta H_0[\vec{x}(\pi,z_0,N)]= \sqrt{\frac{1}{N}\sum_{i=0}^{N}\left(H_0(z_i,\phi_i)^2-\expval{H_0}^2 \right)},
\end{equation}
where $\expval{H_0}=\frac{1}{N}\sum_{i=0}^{N}H_0(z_i,\phi_i)$ is the average of the energy along the trajectory. This standard deviation  is plotted   as a function of $z$  in Fig. \ref{fig: appendix} Panel (b). The discontinuity observed in the standard deviation indicates the location of the separatrix of the island. The  two distinct branches of the separatrix are shown in   red and green  in Fig.~\ref{fig: appendix}, Panel (a). The areas below these lines are denoted, respectively,  as $S_{r:s}^{+}$ (red) and $S_{r:s}^{-}$ (green). The difference  $S_{r:s}^{+}-S_{r:s}^{-}$ represents the phase-space area enclosed between  the two separatrix branches. Using the same procedure, the stable fixed point can also be identified, as the point where $\Delta H_0=0$ (fixed point), which is again indicated with a solid black dot. 

Now, to calculate the trace of $M_{r:s}$, we first identify the stable fixed point and then compute the monodromy matrix
\begin{equation}
    M_{r:s}=\begin{pmatrix}
    \pdv{\phi(r\tau)}{\phi(0)} & \pdv{\phi(r\tau)}{z(0)} \\[6pt]
    \pdv{z(r\tau)}{\phi(0)} & \pdv{z(r\tau)}{z(0)}
    \end{pmatrix},
\end{equation}
after $r$ periods, where the derivatives are approximated numerically. To validate the results, we check that $\tr M_{r:s}<2$ (regular point) and $\det M_{r:s}=1$ (unitarity).

\subsection{Quasi-energy splitting for the kicked LMG model}
\begin{figure*}
    \centering
    \includegraphics[width=\linewidth]{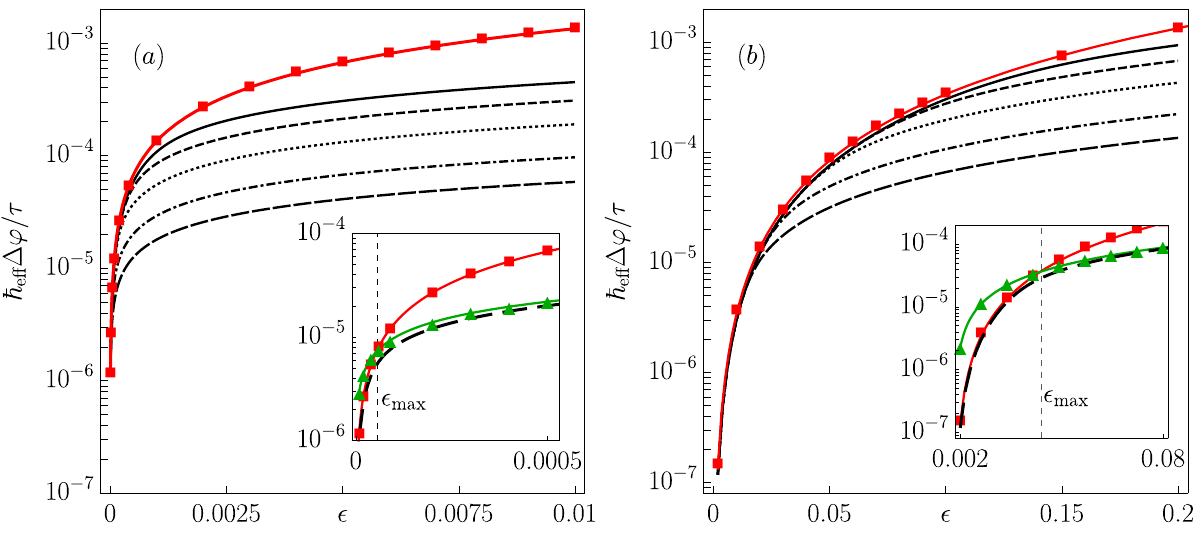}
    \caption{\justifying{Eigenphase difference (scaled with $\hbar_{\text{eff}}$) as a function of $\epsilon$, the perturbation strength. Panel (a) shows a \( 1{:}1 \) resonance. Panel (b) shows a \( 2{:}1 \) resonance. In all cases, $J=60$ (solid line), $J=90$ (dashed line), $J=150$ (dotted line), $J=300$ (dashed-dotted line) and $J=500$ (long dashed line). Red squares shows the semiclassical prediction  given by \eqref{eq: quasienergy splitting}. Inset with a zoom in the nearly integrable regime is included, where the green triangles shows the Harmonic regime \eqref{eq: quasienergy splitting 2} for $J=300$.}}
    \label{fig: RAT regime}
\end{figure*}

The quasienergy splitting (scaled with $\hbar_\text{eff}$) is presented in Fig.~\ref{fig: RAT regime} for the two resonances under consideration. Panel (a) corresponds to the $1{:}1$ resonance, where results for several values of the effective Planck constant are shown, while panel (b) displays the analogous situation for the $2{:}1$ resonance. In both cases, the red squares represent the semiclassical prediction obtained from RAT theory, Eq. \eqref{eq: quasienergy splitting}, whereas the black solid lines correspond to the exact quantum results, computed through the diagonalization of the Floquet matrix. 

From these plots, two clearly distinct regimes can be identified. In the first regime, corresponding to relatively small values of the perturbation strength, the quantum results exhibit excellent agreement with the semiclassical prediction for each system size (i.e., for each value of $J$) considered. This agreement confirms the validity of RAT theory in this parameter region. In the second regime, occurring as the perturbation strength increases, the quantum results no longer follow the semiclassical curve but instead saturate, indicating the breakdown of RAT theory. Interestingly, the semiclassical approximation works better in the quantum regime with smaller $J$, a situation that may seem paradoxical, but in reality is related with the size of the classical structures in comparison with the effective Planck constant. 

Similar results have been reported in previous studies of other systems such as the Standard and Harper maps  \cite{Wisniacki_2014, Wisniacki_2015}. In those works, it was conjectured that the observed saturation of the quasienergy splitting is a consequence of the two resonant states becoming fully quantized within the resonance islands. Once this occurs, their splitting is no longer governed by RAT theory but instead corresponds to that of a harmonic oscillator, whose frequency is given by the local oscillation frequency associated with the resonance. We verify this explicitly in the insets of each panel. The insets provide a zoom into the nearly integrable regime, where the semiclassical RAT prediction is again plotted with red squares. For the sake of clarity, only the exact quantum results for $J=300$ are displayed. The harmonic oscillator approximation derived in Eq.~\eqref{eq: quasienergy splitting 2} is shown with solid green triangles. The comparison corroborates that, as $\epsilon$ increases and the resonance islands expand in phase space, the quantum states effectively become the eigenstates of a harmonic oscillator localized within the resonance. Although only one value of $J$ is plotted in the insets for better visibility, we have verified that the same behavior holds consistently for all the effective Planck constants  $\hbar_\text{eff}=1/J$ considered (not shown).  In each inset we additionally include a vertical gridline indicating an upper bound for the perturbation strength, denoted as $\epsilon_{r{:}s}^{\text{max}}$, which marks the transition between the RAT regime to the harmonic-oscillator regime. Beyond this point, RAT theory can no longer be considered reliable, and it is replaced by the harmonic approximation. This upper bound is determined as the intersection between the Harmonic and RAT curves, and depends on both the system size $J$, and the resonance order $r$. In the next subsection, we employ both analytical expressions and numerical analysis to determine the scalling behaviour of $\epsilon_{r{:}s}^{\text{max}}$. 
\subsection{Transition to the Harmonic regime}
From the insets in Fig.~\ref{fig: RAT regime}, we can observe that there is an intersection between the RAT and the harmonic approximations, that is, a value of $\epsilon$ where both approximations predict the same quasienergy splitting. We define this intersection as $\epsilon_{r{:}s}^{\text{max}}$, since it marks the point beyond which the RAT approximation is no longer valid and must be replaced by the harmonic one. The intersection is obtained by equating Eqs.~\eqref{eq: quasienergy splitting} and \eqref{eq: quasienergy splitting 2}, we obtain
\begin{equation}
    \label{eq: equality}
    2\abs{K_{r:s}}=\frac{r^2\hbar_{\text{eff}}^2}{m_{r:s}}.
\end{equation}
Because $K_{r:s}$ depends on the size of the resonance island, we can write $K_{r:s}=K_{r:s}(\epsilon)$. From Eq.~\eqref{eq: equality}, the value of $\epsilon_{r{:}s}^{\text{max}}$ can in principle be determined. Unfortunately, the explicit dependence of $K_{r:s}$ on $\epsilon$ cannot be derived analytically. However, it can be inferred numerically, allowing us to extract the scaling behavior. 

\begin{figure}
    \centering
    \includegraphics[width=\linewidth]{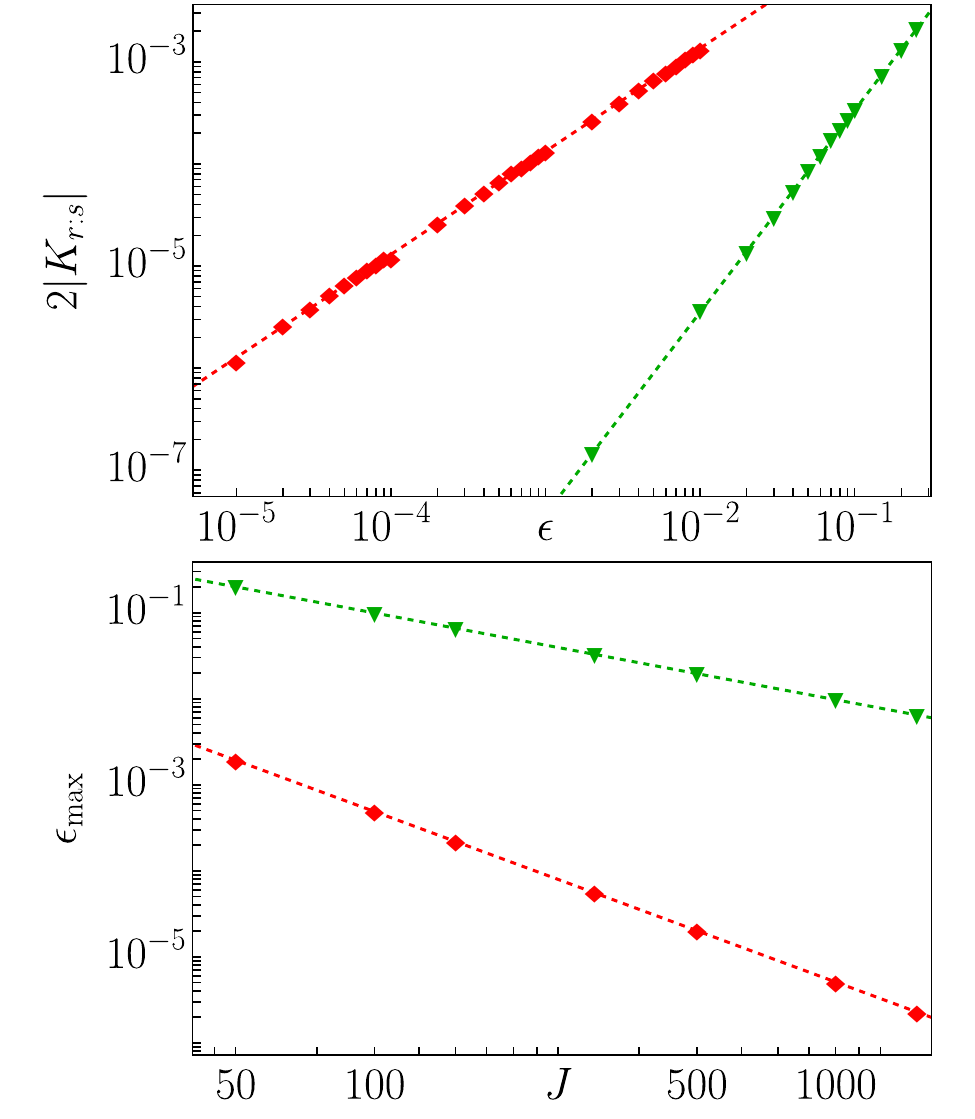}
    \caption{\justifying{ (a) Scaling of $2\abs{K_{r:s}}$ with respect to $\epsilon$. (b) Scaling of $\epsilon_{r{:}s}^{\text{max}}$ with respect to $J$. In all cases, log-log scale is used. Solid red diamonds corresponds to $1{:}1$  resonance and solid green triangles to $2{:}1$  resonance. We have included the fitting functions with dashed lines (see text for details).}}
    \label{fig: scalling}
\end{figure}

In Fig.~\ref{fig: scalling}, we plot in log-log scale the parameter $2\abs{K_{r:s}}$ for the two resonances considered. The red diamonds correspond to the $1{:}1$ resonance, while the green triangles show the $2{:}1$ case. The dashed lines represent numerical fits, which in both cases reveal a linear growth but with distinct slopes. The resulting fits are $\log(K_{1:1}(\epsilon))=-1.9525+1.00979\log\epsilon$ and $\log(K_{2:1}(\epsilon))=-3.4325+1.98108\log\epsilon$, respectively. This power-law behavior its in fact inherited from the growth of the resonance island area $(S^{+}_{r{:}s} - S^{-}_{r{:}s})$, since $K_{r:s}$ is proportional to it. 

In contrast, the effective mass parameter $m_{r:s}$ remains constant with $\epsilon$. Therefore, both the quasienergy splitting and $\epsilon_{r{:}s}^{\text{max}}$ depend solely on the island size. From Eq.~\eqref{eq: equality}, we conclude that $\epsilon_{r:s}^\text{max}$ goes to zero at different rates depending on the resonance 
\[
\epsilon^{\text{max}}_{1:1}\propto\hbar_{\text{eff}}^2=\frac{1}{J^2}, \qquad \epsilon^{\text{max}}_{2:1}\propto\hbar_{\text{eff}}=\frac{1}{J}.
\]

This analytical prediction is confirmed numerically in Fig.~\ref{fig: scalling}, where we show $\epsilon^{\text{max}}_{r:s}$ as a function of $J$ for both resonances. The fits obtained are $\log(\epsilon^{\text{max}}_{1:1})=1.50101-1.98149\log J$ and $\log(\epsilon^{\text{max}}_{2:1})=2.33651-1.00856\log J$, in agreement with the expected scaling. Thus, the transition between the RAT and harmonic regimes crucially depends on the integers $r{:}s$, since the growth rate of the resonance islands differs for each case.

This scaling behavior is directly related to the growth rate of the phase-space area occupied by the resonances: for a $1{:}1$ resonance, the area grows as $\sqrt{\epsilon}$, whereas for a $2{:}1$ resonance, it grows linearly with $\epsilon$. Consequently, for a $1{:}1$ resonance, the area reaches the value of $\hbar_{\text{eff}}$ as $\epsilon\propto J^{-2}$, while for a $2{:}1$ resonance as $\epsilon\propto J^{-1}$. Interestingly, the differing growth rates of the resonance islands for the 1:1 and 2:1 resonances can be understood through a parity selection rule that emerges in the quantum version of the model. The perturbation used in this study changes sign under the transformation $\hat{\Pi}\hat{J}_x \hat{\Pi}^\dagger=-\hat{J}_x$.  As previously mentioned, for the parameters chosen in this study, the eigenstates of $\hat{H}_0$ exhibit alternating parity eigenvalues  ($\pm$).  Consequently, for the 1:1 resonance, which involve consecutive  energy levels, the  matrix element    $\bra{E_{k_{R}+1}}\hat{J}_x \ket{E_{k_R}}\not=0$, making the perturbation effective.  In contrast,  the  2:1 resonance involves     next-to-consecutive levels that share   the same parity,  resulting in vanishing matrix elements, $\bra{E_{k_R+2}}\hat{J}_x \ket{E_{k_R}}=0$, and thus a reduced perturbative effect.

\section{CONCLUSIONS}
In this work, we have analyzed the validity of resonance-assisted tunneling (RAT) theory in a quantum kicked many-body system. By means of a quantum resonance condition analogous to its classical counterpart, we were able to identify resonant states within the quantum spectrum. Our results show that RAT theory remains applicable for resonant states up to a maximum perturbation strength, $\epsilon_{r{:}s}^{\text{max}}$, which depends on the effective Planck constant and vanishes in the semiclassical limit. This threshold coincides with the point at which resonant states become fully quantized within the resonance islands and scales as $\hbar_{\text{eff}}^2$ for a $1{:}1$ resonance, and as $\hbar_{\text{eff}}$ for a $2{:}1$ resonance. \\

We additionally show that, beyond the RAT validity regime, the quantum states transform into eigenstates of an effective harmonic oscillator, with a frequency determined by the local oscillation frequency of the resonance. Our numerical results show excellent agreement with the analytical predictions. The intermediate regime, in which the eigenstates transition from the RAT regime to the harmonic regime is far from obvious, and further efforts can be made in this direction. 
\section{ACKNOWLEDGMENTS}
We acknowledge the support of the Computation Center - ICN, in particular to Enrique Palacios, Luciano D\'iaz, and Eduardo Murrieta. JAS-L and J.G.H acknowledge partial financial support from the DGAPA-UNAM project IN109523.
JAS-L is grateful to the Secretar\'ia de Ciencias, Humanidades, Tecnolog\'ia e Innovaci\'on (SECIHTI) for founding his graduate education,  CVU number:1181841. D.A.W.  received support from CONICET (Grant No. PIP 11220200100568CO), UBACyT (Grant No. 20020220300049BA), ANCyPT (Grants  No. PICT2020-SERIEA-01082) and PREI-UNAM.

\appendix
\section{Quantum periods}
\label{App:QP}
As discussed in the main text, we adjusted the period of the quantum system in order to exactly fulfill the resonance condition for each value of $\hbar_{\text{eff}}$ considered. In Refs.~\cite{seguralanda2025,segura2024quantum}, it was demonstrated that the quantum period \eqref{eq: quantum periods}, when defined using nearest neighbors, converges to the classical period in the limit $\hbar_{\text{eff}}\to0$. The purpose of this appendix is to compare the exact quantum periods given by Eq.~\eqref{eq: quantum periods} for each resonance under study. Classically, the $1{:}1$ resonance corresponds to $\tau=8$, whereas the $2{:}1$ resonance is obtained with $\tau=4$.  

In the following table, we report the quantum periods computed from the exact eigenvalues of $\hat{H}_0$. The first column lists the values of $J$ considered. The second and third columns display the quantum periods obtained using nearest- and second-neighbor levels, respectively, where the index $k_R$ denotes the eigenstate whose energy is closest to the classical resonant energy $E/J=-0.723276$. As expected, the quantum period approaches the classical value as $J$ increases. Even for the smallest value considered ($J=30$), the difference remains relatively small. Consequently, the classical information required to approximate the quantum splitting is essentially unaffected by these minor variations in $\tau$. In particular, the phase-space area of the resonance island is not significantly altered by such fluctuations, thereby justifying the use of a fixed classical period while allowing the quantum one to vary slightly.

\begin{table}[h]
\centering
\begin{tabular}{c c c}
\toprule
$J$  & $\Big(T_{k_R+1,k_R}\Big)$   & $\Big(T_{{k_R+2},k_R}\Big)$   \\ 
\specialrule{1.2pt}{2pt}{2pt} 
30   & 7.90344990884333  & 4.003373105555204 \\
60   & 7.956160393204096 & 4.003979081610065 \\
90   & 7.973927688513776 & 4.004247984195059 \\
150  & 7.988216500195171 & 4.004487842190517 \\
300  & 7.998978076540702 & 4.004682359106553 \\
500  & 7.995012508618674 & 4.000612049866183 \\
1000 & 7.998243854681983 & 4.000675148524783 \\
\bottomrule
\end{tabular}
\caption{Exact values of the quantum periods for each $J$ considered.}
\end{table}

\bibliography{references}
\end{document}